\documentclass[aps,pra,superscriptaddress,twocolumn,floatfix,10pt]{revtex4-1}

\usepackage{amsmath}
\usepackage{bigints}
\usepackage[draft]{changes}
\usepackage{graphicx}
\usepackage{graphicx,amsmath,amssymb,amsthm,amsfonts}
\usepackage[utf8x]{inputenc}
\usepackage{url}
\usepackage{float}
\usepackage{xcolor}
\usepackage{hyperref}
\usepackage{lipsum}
\usepackage{bm}
\usepackage{soul}
\usepackage{comment}
\usepackage[many]{tcolorbox} 
\usepackage{cleveref}

\def\r{{\mathbf{r}}}

\def\eps{{\varepsilon}}
\def\E{{\mathbf{\widetilde{E}}}}
\def\H{{\mathbf{\widetilde{H}}}}
\def\B{{\mathbf{\widetilde{B}}}}

\def\tomega{{\tilde{\omega}}}

\definechangesauthor[name={Kevin Vynck}, color=orange]{KV}
\definechangesauthor[name=<Ashod>, color=teal]{AA}
\begin{document}

\title{Symmetry-Engineered Magnetic Dipole Emission in Plasmonic Core–Satellite Resonators}

\author{Joshua Davis}
\affiliation{Univ. Bordeaux, CNRS, CRPP, UMR 5031, F-33600 Pessac, France}

\author{Sébastien Bidault}
\affiliation{Institut Langevin, ESPCI Paris, Université PSL, CNRS, F-75005 Paris, France}

\author{Mathieu Mivelle}
\affiliation{Sorbonne Université, CNRS, Institut des Nanosciences de Paris, INSP, F-75005 Paris, France}

\author{Mona Tréguer-Delapierre}
\affiliation{Univ. Bordeaux, CNRS, Bordeaux INP, ICMCB, UMR 5026, F-33600 Pessac, France}

\author{Alexandre Baron}
\email{alexandre.baron@u-bordeaux.fr}
\affiliation{Univ. Bordeaux, CNRS, CRPP, UMR 5031, F-33600 Pessac, France}
\affiliation{Institut Universitaire de France, 1 rue Descartes, 75231 Paris Cedex 05, France}

\date{\today}

\begin{abstract}
\noindent Magnetic dipole (MD) transitions are intrinsically weak and highly sensitive to emitter orientation and position, making their controlled enhancement at optical frequencies particularly challenging. Here we show that structural symmetry provides a powerful route to robust magnetic light–matter interactions. We systematically investigate plasmonic core–satellite resonators composed of $N$ metallic nanoparticles arranged on a dielectric core. We evaluate their performance using a unified figure of merit that accounts for magnetic Purcell enhancement, electric dipole suppression, quantum efficiency, and robustness to emitter orientation and fabrication tolerances. We find that the optimal structures correspond to the highest-symmetry geometries, which naturally produce spatially homogeneous and orientation-independent magnetic Purcell enhancement. In particular, the dodecapod configuration yields strong magnetic emission with Purcell factors approaching 590, good radiative efficiency, and suppressed electric dipole contributions. Quasinormal-mode and complex mode-volume analysis reveal that symmetry enforces uniform magnetic modal confinement within the core, explaining both the enhancement and its robustness. These results establish symmetry as a guiding principle for designing nanophotonic resonators with controlled multipolar light–matter interactions and provide a practical route toward bright and selective magnetic dipole emitters.
\end{abstract}

\maketitle

\noindent Light–matter interactions are most commonly described in terms of electric dipole (ED) transitions coupled to the electric field of light. More generally, the interaction Hamiltonian contains higher-order multipolar contributions,
$
H_{\mathrm{int}} = -\mathbf{p}\cdot\mathbf{E} - \mathbf{m}\cdot\mathbf{B} - \dots,
$
where $\mathbf{p}$ and $\mathbf{m}$ denote the electric and magnetic transition dipole moments, and $\mathbf{E}$ and $\mathbf{B}$ are the electric and magnetic fields~\cite{Sanz-Paz2006}. In certain systems, such as transition-metal ions (e.g.\ $\mathrm{Cr^{3+}}$)~\cite{Karaveli2013} and lanthanide emitters including $\mathrm{Eu^{3+}}$ and $\mathrm{Er^{3+}}$~\cite{Vaskin2019,Choi2016}, ED transitions are parity forbidden and magnetic dipole (MD) emission becomes non-negligible. In practice, because of ligand-field effects, the ED emission remains dominant~\cite{Cowan1981}. The design of efficient lanthanide-based emitters has relied extensively on chemical engineering to maximize the forced electric dipole transitions using low-symmetry ligand fields, overlooking the allowed magnetic transitions~\cite{de2000spectroscopic}. Engineering photonic environments capable of selectively enhancing MD emission while suppressing competing ED channels therefore remains a central challenge.

Since Purcell’s seminal work~\cite{Purcell1946}, it is well known that the spontaneous emission rate of a quantum emitter can be modified by tailoring its electromagnetic environment. Plasmonic and dielectric nanostructures can produce strong field confinement and large Purcell factors, enabling dramatic control of spontaneous emission. Numerous resonator geometries have therefore been proposed to enhance MD transitions, including split-ring resonators, dielectric Mie resonators, and other engineered nanocavities~\cite{Hein2013,feng2017ideal,Feng2018,Sugimoto2021,reynier2025nearfield}. While large theoretical enhancements have been predicted, these designs often suffer from severe practical limitations: the Purcell enhancement typically depends strongly on the orientation and precise position of the emitter, and the largest predicted effects frequently occur only in relatively large or complex structures~\cite{Wu2019,brule2022magnetic,Puente2023,utyushev2024generation}. These constraints severely limit experimental implementation and highlight the need for nanophotonic architectures that combine strong MD enhancement with robustness to emitter orientation, position, and fabrication variations.

Here we introduce a symmetry-driven design strategy for nanophotonic resonators that enables robust control of multipolar light-matter interactions. The central idea is that specific symmetric geometries naturally suppress anisotropic responses and can produce orientation-independent electromagnetic environments for emitters. Magnetic dipole emission provides a stringent test case for this concept, as it requires the simultaneous optimization of several competing quantities, including magnetic Purcell enhancement, suppression of electric dipole emission, high quantum efficiency, and robustness to emitter orientation and position. Previously, symmetry analysis has been introduced for the multipole classification of isolated resonators,~\cite{xiong2020, gladyshev2020}, yet the general implementation for resonators consisting of multiple particles, and the implications for control of the isotropy of the local electromagnetic density of states remain unexplored.

\begin{figure*}[t!]
	\centering	
    \includegraphics[width=\textwidth]{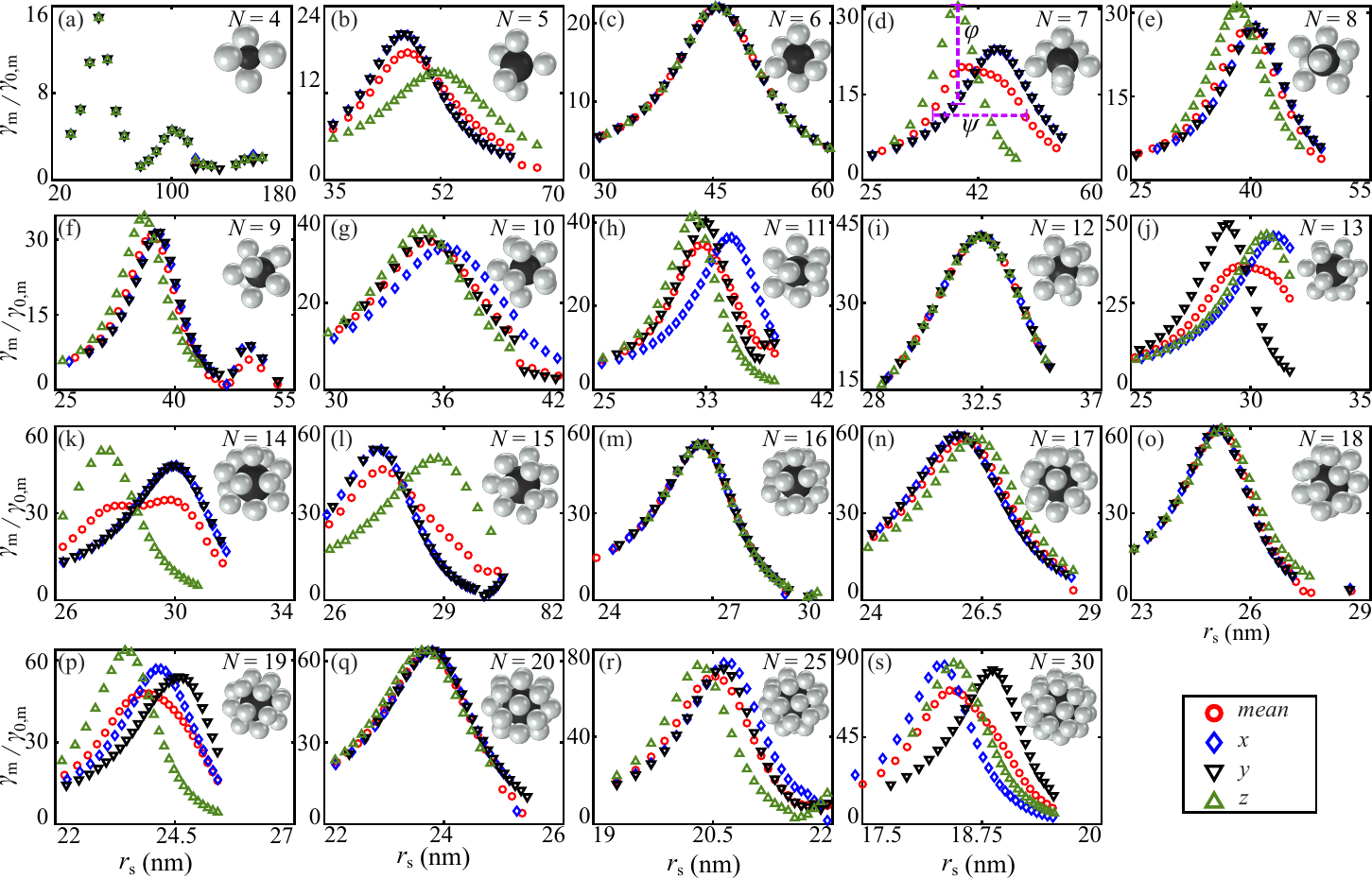}
	\caption{Magnetic Purcell factor as a function of the satellite radius $r_\mathrm{s}$ for the plasmonic $\mathit{N}$-pod resonators considered in this work. Insets show the corresponding particle geometries, consisting of a silica core of radius $r_\mathrm{c}=42.5$~nm (dark gray, $n = 1.46$ over the desired spectral range) surrounded by silver satellites (light gray). Red circles denote the orientation-averaged magnetic Purcell factor, while blue diamonds, black inverted triangles, and green triangles correspond to magnetic dipoles oriented along the $x$, $y$, and $z$ directions, respectively.}
	\label{Fig:orientation_image}
\end{figure*}
As a concrete implementation of this symmetry-engineering approach, we investigate plasmonic multipod structures consisting of $N$ identical metallic nanoparticles symmetrically arranged around a dielectric core containing the emitter. Such core–satellite architectures have already been realized experimentally through colloidal nano-chemistry~\cite{Thill2012,many2019high,lermusiaux2021toward, roach2023symmetric}. 
Using a quasi-normal-mode framework and a unified figure of merit, we show that tetrahedral structural symmetry leads to robust and isotropic magnetic dipole enhancement, revealing symmetry as a key ingredient for robust magnetic dipole emission. Specific high-symmetry geometries, particularly the dodecapod configuration, generate strong and spatially homogeneous magnetic Purcell enhancement while simultaneously suppressing electric dipole emission. The resulting structures provide bright, efficient, and orientation-independent magnetic light sources with Purcell factors approaching 590, demonstrating the power of symmetry as a guiding principle for nanophotonic resonator design. Remarkably, we observe spatially homogeneous fields over the entire dielectric core, with very little perturbation by the near field of the plasmonic satellites.

Designing complex nanostructures for MD enhancement remains challenging because the parameter space is often vast. For simple geometries, analytical or semi-analytical techniques are effective~\cite{Moroz2005}. For arbitrarily shaped structures, numerical studies typically rely on time-domain methods like finite-difference time-domain (FDTD)~\cite{Taflove2005} or direct finite element methods (FEM)~\cite{Monk2003}. While robust, these approaches require repeated simulations for each source location, orientation, and wavelength on top of variations in the numerous geometrical parameters, making optimization over large parameter spaces computationally expensive. Moreover, they often obscure the underlying modal physics of the resonator. An emerging alternative is the quasi-normal mode (QNM) framework~\cite{Lalanne2018}, in which the behavior of the nanostructure is described in terms of its natural resonances. Each QNM is a solution to the source-free Maxwell equations and is characterized by a complex eigenfrequency $\tomega$. Once the QNMs of a structure are known, physical observables such as Purcell factors can be computed efficiently via overlap integrals, eliminating the need to scan over source configurations and frequencies. This modal approach not only accelerates optimization but also provides direct physical insight into the mechanisms and symmetry driving MD enhancement.

We consider plasmonic $\mathit{N}$-pods (see insets of Fig.~\ref{Fig:orientation_image}) consisting of $\mathit{N}$ identical silver nanospheres of radius $r_\mathrm{s}$ symmetrically arranged around a silica core of radius $r_\mathrm{c}=42.5$~nm, representative of experimentally realized systems~\cite{lermusiaux2021toward}. The maximum satellite size is chosen to prevent inter-particle contact. The silver permittivity is modeled using a single-pole Drude-Lorentz model fitted to the data of Johnson and Christy~\cite{johnson1972optical}. Satellite positions are determined by solving the Thomson problem~\cite{tomson1904structure}, yielding exact solutions for $\mathit{N}=\{1,2,3,4,5,6,12\}$ and numerically optimized configurations otherwise~\cite{Wales2006}. Each structure has a characteristic polyhedron associated with it for which the face centers correspond to the satellite centers (see the online supplemental material for a list of the equivalent polyhedra and associated symmetry classifications)~\cite{note:SM}. The $N$-pods inherit symmetry properties from the associated polyhedra.
We focus on magnetic and electric dipole transitions at $\lambda_\mathrm{m}=590$~nm and $\lambda_\mathrm{e}=610$~nm, respectively, corresponding to $\mathrm{Eu^{3+}}$ emission lines~\cite{Vaskin2019}. The PF for a magnetic dipole emitter located at $\mathbf{r}_0$ is expressed within the quasi-normal mode (QNM) framework as
\begin{equation}
\label{eq:Purcell_factor_mag}
\frac{\gamma_\mathrm{m}(\omega)}{\gamma_{0,\mathrm{m}}(\omega)}
=
\frac{2}{\hbar\gamma_{0,\mathrm{m}}(\omega)}
\sum_j
\mathrm{Im}\left[\alpha_{j,\mathrm{m}}(\omega)\mathbf{m}^*(\omega).\B_j(\mathbf{r}_0,\tomega_j)
\right],
\end{equation}
where $\gamma_\mathrm{m}$ is the magnetic decay rate in the presence of the nanostructure, $\gamma_{0,\mathrm{m}}$ is the total decay rate without the nanostructure, $\textbf{m}$ is the magnetic dipole moment, $\B_j$ is the magnetic field of the $j$th QNM with complex eigenfrequency $\tilde{\omega}_j$ and $\alpha_{j,\mathrm{m}}(\omega)=-\omega(\tilde{\omega}_j-\omega)^{-1}\![\mathbf{m}\!\cdot\!\B_j(\mathbf{r}_0,\tomega_j)]$ is the corresponding excitation coefficient. A similar formula is defined for the PF of an electric dipole emitter, by replacing all subscripts `m' with `e', $\mathbf{m}$ with $\mathbf{p}$, $\B_j$ with $\E_j$ and where $\alpha_{j,\mathrm{e}}(\omega)=\omega(\tilde{\omega}_j-\omega)^{-1}\![\mathbf{p}\!\cdot\!\E_j(\mathbf{r}_0,\tomega_j)]$. 
\begin{figure}[t!]
	\centering	
    \includegraphics[width=0.3\textwidth]{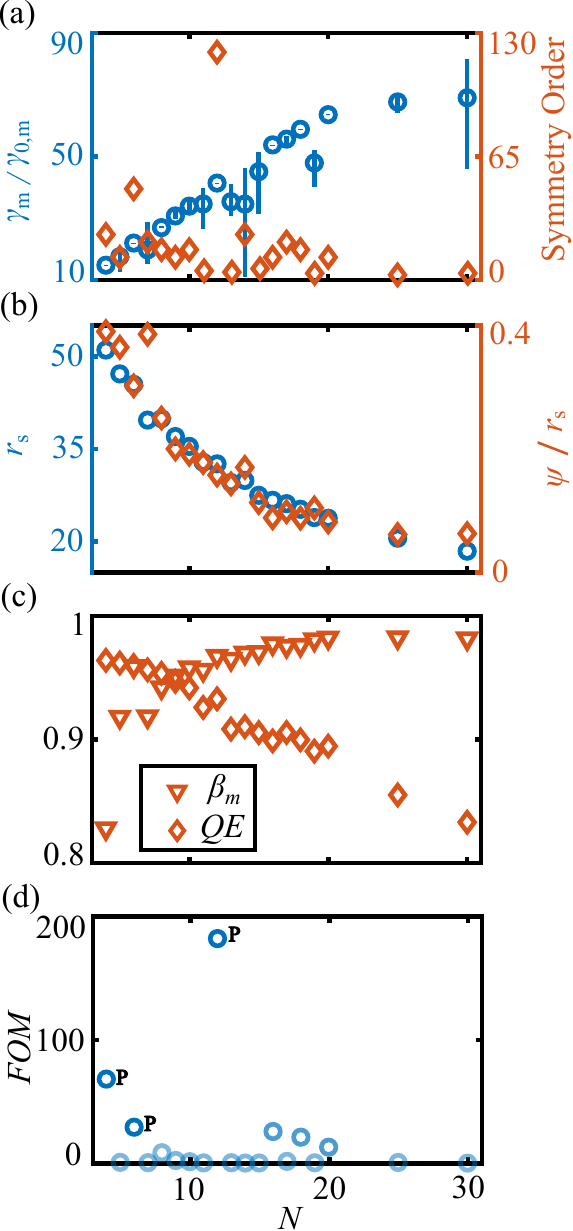}
	\caption{All plots show the dependence of various quantities with the number $N$ of satellites in the plasmonic core-satellite structure. (a) The left  axis and blue dots are the maximum value of the average magnetic PF, the blue error bars show $\varphi$ the maximum spread of the $x,y,z$ oriented PFs at the same radius, the orange diamonds and right axis show the symmetry order of the equivalent polyhedra.  (b) The blue dots and left axis are the optimal satellite radius of the averaged PF, while the orange diamonds and right axis are the full width half maximum values of the peak normalized to $r_s$. (c)  Branching ratio $\beta_m$ (inverted triangles) and quantum efficiency QE (diamonds). (d) $FOM$ calculated with \cref{eq:FOM}, the ``P" next to the dots indicate a particle with a Platonic equivalent polyhedron. }
	\label{Fig:summaries}
\end{figure}

All simulations are performed with the particles immersed in water, which provides a realistic environment for colloidal suspensions, although larger Purcell factors could be achieved in air due to stronger refractive-index contrast. Electromagnetic fields are computed using the finite-element method implemented in COMSOL Multiphysics, and QNMs are extracted and normalized using the MAN freeware \cite{Wu2023}. The minimum separation between the satellites and the core is fixed to 2~nm to ensure reliable meshing. Each $\mathit{N}$-pod is excited by a magnetic dipole placed at the center of the core and oriented along three orthogonal directions, and the magnetic Purcell factor is computed as a function of the satellite radius $r_\mathrm{s}$ (see Fig.~\ref{Fig:orientation_image}). For each $\mathit{N}$, we extract the maximum orientation-averaged Purcell factor and its spread $\varphi$, defined as the difference between the maximum and minimum values over dipole orientations at the optimal $r_\mathrm{s}$ for the average Purcell (see Fig. \ref{Fig:orientation_image}(d) for an illustration of $\varphi$).

Figure~\ref{Fig:summaries}(a) shows that the maximal averaged Purcell factor initially increases with $\mathit{N}$ and saturates for $\mathit{N}\gtrsim20$. Figure \ref{Fig:orientation_image} shows that a clear correlation emerges between symmetry and orientation robustness. $\mathit{N}$-pods exhibiting tetrahedral symmetry ($N=\{4,6,12,16\}$) - including all Platonic solids ($N=\{4,6,12\}$) - display nearly complete independence from dipole orientation. This is because, in the case of a resonator exhibiting tetrahedral symmetry, the isotropy of the magnetic PF is verified, as shown analytically in section II of the online supplemental material \cite{note:SM}. In contrast, low-symmetry structures show strong orientation dependence ($N=\{11,13,15,19,25,30\}$), while particles with high symmetry around a single rotational axis are insensitive to two dipole orientations but remain strongly dependent on the third ($N=\{5,7,14\}$). These trends directly reflect the spatial distribution of the satellites (Fig.~\ref{Fig:orientation_image}) and demonstrate that tetrahedral symmetry is optimal for achieving large, orientation-independent magnetic Purcell enhancement. Importantly, the orientation dependence of the structures shown in Fig. \ref{Fig:orientation_image} is refelcted identically in the degeneracy of the dominant MD QNMs, (see Section II of the online supplimentary material for a full discussion and table of important modes, \cite{note:SM}). For example, all structures which show magnetic PF isotropty for three orientations have a triply degenerate MD mode, while those with isotropy in two dimensions have two of the three MD modes being degenerate. The optimal satellite radius of the averaged Purcell factor decreases with $N$, which is expected since packing more non-intersecting spheres on the core of fixed size requires reducing their radius. The full width at half maximum (FWHM), denoted by $\psi$ and illustrated in Fig.~\ref{Fig:orientation_image}(d), of the orientation-averaged Purcell factors is plotted in Fig.~\ref{Fig:summaries}(b). The linewidth decreases rapidly with increasing $\mathit{N}$, indicating a growing sensitivity to the satellite radius. For instance, at $\mathit{N}=20$, a deviation of only 1~nm from the optimal $r_\mathrm{s}$ reduces the Purcell factor by approximately $50\%$. Such stringent fabrication tolerances are challenging to achieve experimentally, suggesting that despite their larger peak enhancements, high-$\mathit{N}$ structures may be impractical for robust implementations.

Having identified the global maxima of the orientation-averaged magnetic Purcell factor for each particle, we now assess the quality of the $\mathit{N}$-pods as magnetic dipole sources. To this end, we introduce a figure of merit (FOM) that captures source robustness and performance within a single metric. The FOM incorporates spatial robustness through $\bar{\gamma}_\mathrm{m}$, defined as the volume-averaged magnetic dipole PF, where the averaging is performed over the emitter position $\mathbf{r}_0$ inside the core. To calculate $\bar{\gamma}_m$, we placed dipole sources at $93$ evenly sampled locations inside the core and computed the orientation averaged PF at each location, from which the volume mean and standard deviation are calculated, (see appendix B and Figure 3 of the online SM \cite{note:SM} for more information). Remarkably, in all cases the standard deviation is less than $10\%$ of the mean PF. For the dodecapod this decreases to $2\%$. Source purity is quantified by the branching ratio $\beta_\mathrm{m}=\gamma_\mathrm{m}/\gamma$, where $\gamma = \gamma_\mathrm{m}+\gamma_\mathrm{e}$ is the total decay rate of the dressed emitter. Radiative efficiency is characterized by the quantum efficiency, which we define as $QE=\gamma_\mathrm{rad}/\gamma$, where $\gamma_\mathrm{rad}=\gamma-\gamma_\mathrm{NR}$, where $\gamma_\mathrm{NR}$ is the non-radiative decay rate. It is computed through a modified expression from \cite{Sauvan2013}, to include multiple modes 
\begin{equation}
    \gamma_\mathrm{NR}=\frac{2}{\hbar}\int \mathrm{Im}\!\left[\varepsilon(\mathbf{r})\right] \left\lvert \sum_j \alpha_j\mathbf{E}_j(\mathbf{r})\right\rvert^2 \mathrm{d}^3\mathbf{r}
\end{equation}
Robustness to polydispersity is accounted for through the normalized linewidth $\psi/r_\mathrm{s}$, while sensitivity to dipole orientation is described by the spread $\varphi$, which should be minimized. The FOM is therefore defined as
\begin{equation}
\label{eq:FOM}
\mathrm{FOM}
=
\frac{\bar{\gamma}_\mathrm{m}}{\varphi}\beta_\mathrm{m}\,QE\frac{\psi}
     {r_\mathrm{s}}.
\end{equation}
$\beta_\mathrm{m}$ and $QE$ are plotted as a function of $\mathit{N}$ in Fig.~\ref{Fig:summaries}(c), while the resulting FOM is shown on Fig.~\ref{Fig:summaries}(d). The spectral distributions of $\gamma_\mathrm{m}/\gamma_0$, $\gamma_\mathrm{e}/\gamma_0$, $\beta$ and $QE$ for every $\mathit{N}$-pod considered are available on the online supplemental material (see Fig. S1)~\cite{note:SM}. Additionally, we note that the choice of the FOM is not unique and reflects a simple way to characterize many different parameters (see section IV of ~\cite{note:SM} for a more complete discussion on the optimization methodolgy).

As shown in Fig.~\ref{Fig:summaries}(c), the $QE$ decreases monotonically with increasing $\mathit{N}$, while the branching ratio $\beta_\mathrm{m}$ rises rapidly and saturates above $\beta_\mathrm{m}\approx0.95$ for $\mathit{N}\gtrsim12$. Notably, all $\mathit{N}$-pods with $5\leq \mathit{N}\leq20$ exhibit both $QE$ and $\beta_\mathrm{m}$ exceeding $88\%$, highlighting their intrinsic robustness. The FOM identifies the three Platonic solids (marked “P” in Fig.~\ref{Fig:summaries}(d)) as optimal candidates, with the dodecapod ($\mathit{N}=12$) maximizing the FOM.

Next, we analyze the complex mode volumes ($\mathcal{V}_\mathrm{E,H}$) of the degenerate MD QNMs, defined as~\cite{Wu_2021}
\begin{equation}
V_\mathrm{A}(\mathbf{r},\mathbf{u}_i)
=
\frac{\int \left[ \E \cdot \frac{\partial \omega \eps}{\partial \omega} \E - \H \cdot \frac{\partial \omega \mu}{\partial \omega} \H \right]\mathrm{d}^3\mathbf{r}}
{2\left[\tilde{\mathbf{A}}(\mathbf{r})\cdot\mathbf{u}_i\right]^2},
\end{equation}
where $\tilde{\mathbf{A}}=\varepsilon^{1/2}(\r)\E$ or $\mu^{1/2}(\r)\H$, such that $V_\mathrm{E}$ and $V_\mathrm{H}$ denote the electric and magnetic mode volumes, respectively and $\mathbf{u}_i$ is the orientation of a small perturber, such as the quantum emitter. The numerator corresponds to the normalization integral of the QNMs. Although the QNM fields $\E$ and $\H$ diverge exponentially in the far field, the mathematical convergence of this integral for arbitrary geometries has now been rigorously established~\cite{Sauvan2022}. We introduce the isotropic mode volume $\mathcal{V}_\mathrm{A}$ such that $\mathcal{V}_A^{-1}(\r)=\sum_l\sum_{i=x,y,z} V^{(l)}_\mathrm{A}(\mathbf{r},\mathbf{u}_i)^{-1}$, where $l$ is the set of degenerate modes. Possesing both the MD and ED modes, we reconstruct the spectrally resolved PFs, which are shown in Section III and Figure 1 of the online SM. This figure, along with Table 2 of the SM, show that for all structures the MD mode is spectrally isolated from the ED mode and any higher order multipole modes which exist. Furthermore, there is no significant overlap between these modes and the MD modes at $590$~nm.

\begin{figure}[t!]
	\centering	
\includegraphics[width=0.5\textwidth]{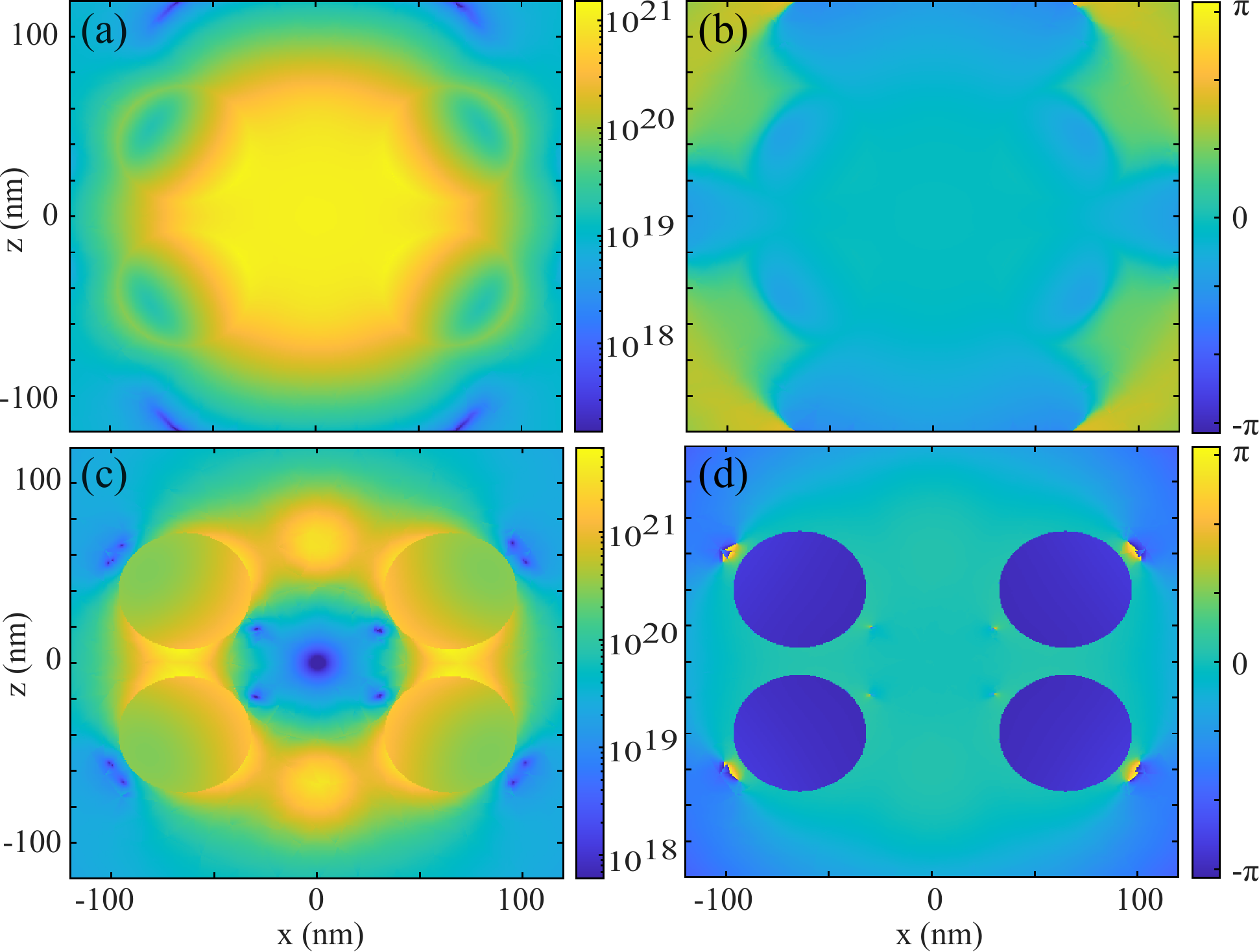}
	\caption{Complex mode volume of the dodecapod consisting of a silica core of radius $r_\mathrm{c}=42.5$~nm surrounded by 12 silver satellites of radius $r_\mathrm{s}=32.5$~nm. (a) Spatial map of the magnitude of the inverse magnetic mode volume $|\mathcal{V}^{-1}_\mathrm{H}(\mathbf{r})|$ (in $\mathrm{m}^{-3}$). (b) Corresponding phase $\arg\!\left(\mathcal{V}^{-1}_\mathrm{H}(\mathbf{r})\right)$. (c) Magnitude of the inverse electric mode volume $|\mathcal{V}^{-1}_\mathrm{E}(\mathbf{r})|$. (d) Corresponding phase $\arg\!\left(\mathcal{V}^{-1}_\mathrm{E}(\mathbf{r})\right)$.}
	\label{Fig:modal_fields}
\end{figure}

The complex isotropic mode volume provides direct insight into the light–matter interaction strength: $|\mathcal{V}_A^{-1}(\r)|$ quantifies the local modal confinement and coupling strength, while $\arg(\mathcal{V}_A^{-1}(\r))$ reflects the non-Hermitian character of the resonance. For a given position $\mathbf{r}$, this phase encodes both the radiation reaction of the cavity mode on a point emitter and the back-action of the emitter on the resonator~\cite{Wu_2021}. As shown in Fig.~\ref{Fig:modal_fields}, the magnetic dipole modes of the dodecapod exhibits a strong and spatially uniform distribution of $|\mathcal{V}^{-1}_\mathrm{H}(\mathbf{r})|$ within the core (Fig.~\ref{Fig:modal_fields}(a)), together with small emitter-cavity dephasing (Fig.~\ref{Fig:modal_fields}(b)). This indicates that magnetic dipole emission is not only strongly enhanced in the core but also robust with respect to emitter positioning. A particularly interesting feature of Fig.~\ref{Fig:modal_fields}, is that there appears to be a negligible effect in the near field region of the plasmonic satellites to both mode volumes inside the core. By contrast, outside the core, near the satellites and in the surrounding medium, $|\mathcal{V}^{-1}_\mathrm{H}(\mathbf{r})|$ rapidly decreases and $\arg\left(\mathcal{V}^{-1}_\mathrm{H}(\mathbf{r})\right)\approx \pi$, indicating an out-of-phase relation between cavity and source and therefore weak magnetic dipole emission. A noticeably different and complementary behavior is observed for the electric dipole mode. The quantity $|\mathcal{V}^{-1}\mathrm{E}(\mathbf{r})|$ is largest in the inter-satellite gaps, albeit with strong spatial inhomogeneity. In this region $\arg\left(\mathcal{V}^{-1}\mathrm{E}(\mathbf{r})\right)$ is also close to 0, implying that an ED emitter placed in the satellite gaps would have a large ED emission. This modal behavior underlies the high FOM of the dodecapod, as the magnetic mode provides volumetric and phase-aligned enhancement within the core, whereas the electric mode volume remains spatially localized outside the core, thereby inhibiting electric dipole emission.

Finally, having established the orientation and location independence of the magnetic PF of the dodecapod, we perform a full optimization with the goal of achieving the highest possible PF. As we have already shown that the dodecapod supports only MD modes in the vicinity of the Eu\textsuperscript{3+} emission wavelength, the parameter space for the optimization is reduced to only two parameters, $r_c$ and $r_s$. In this case direct FEM simulations with a MD source are used at $\lambda = 590$~nm. Once the optimal radii are found, we then use QNM simulations to construct the MD and ED spectra. The satellite and core radii ($r_\mathrm{s}$ and $r_\mathrm{c}$) are optimized using a downhill simplex algorithm combined with basin hopping. The objective function is chosen as $f = \gamma_\mathrm{m}^\mathrm{rad}\,\beta_\mathrm{m}\,QE$, which we maximize in order to design a bright, efficient, and spectrally pure magnetic dipole source. Depending on the desired outcome this cost function can easily be changed, for instance by adding penalty terms to prevent the particle dimensions from becoming too small. 
\begin{figure}[t!]
	\centering	
	\includegraphics[width=0.6\columnwidth]{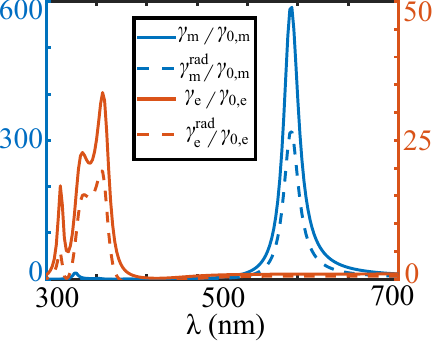}
	\caption{Spectral response of the fully optimized dodecapod with $r_\mathrm{c}=18.5$~nm and $r_\mathrm{s}=19.8$~nm. The solid blue curve (left axis) shows the magnetic Purcell factor, while the solid orange curve (right axis) shows the electric Purcell factor. Dashed curves indicate the corresponding radiative rate enhancements.}
	\label{Fig:full_optimization}
\end{figure}
The spectral response of the optimized structure with dimensions $r_\mathrm{c}=18.5$~nm and $r_\mathrm{s}=19.8$~nm, is shown in Fig.~\ref{Fig:full_optimization}. At the $\mathrm{Eu^{3+}}$ transition wavelength ($\lambda=590$~nm), a magnetic Purcell factor approaching 590 is achieved, with a radiative decay rate enhancement of approximately 290. At the same wavelength, the electric dipole emission is is only weakly enhanced, with a Purcell factor below 2. The electric dipole resonance instead peaks near 400~nm with a Purcell factor of approximately 35. 

In summary, we have systematically investigated core–satellite $\mathit{N}$-pod resonators as platforms for enhancing magnetic dipole emission. By combining finite-element simulations with a quasinormal mode analysis, we demonstrated that geometric symmetry plays a central role in achieving robust and orientation-independent magnetic Purcell enhancement. In particular, the polyhedra associated to the $\mathit{N}$-pods possessing tetrahedral symmetry exhibit near-complete invariance with respect to dipole orientation and position within the core.
To quantitatively compare the different geometries, we introduced a figure of merit that incorporates magnetic transition rate enhancement, branching ratio, quantum efficiency, robustness to polydispersity, and insensitivity to dipole orientation. This unified metric identifies the dodecapod as the optimal compromise between performance and fabrication feasibility. The complex mode volume analysis further reveals that the magnetic mode exhibits strong and spatially uniform confinement within the core, while the electric mode remains localized near the satellites, explaining both the large magnetic enhancement and the suppression of electric dipole emission.

Direct FEM simulations at the $\mathrm{Eu^{3+}}$ transition wavelength demonstrates magnetic Purcell factors exceeding 500 together with good radiative efficiency and weak electric dipole enhancement. These results establish symmetric core–satellite resonators as a realistic and robust route toward bright, spectrally pure magnetic dipole sources. More broadly, our work highlights how modal symmetry engineering can be leveraged to selectively enhance weak optical transitions in nanoscale emitters. While we focus here on magnetic dipole emission, the symmetry-driven design strategy introduced here is general and can be extended to other multipolar and directional light–matter interactions.

\vspace{1cm}

\noindent \textit{Acknowledgements} - This work was supported by the Agence Nationale de la Recherche (ANR-22-CE09-0027). A.B. and J. D. also acknowledges financial support from the Intedisclipnary and Exploratory Research program at University of Bordeaux (RIE HOMOGENIZE). 

\bibliographystyle{ieeetr}
\bibliography{references}

@article{tomson1904structure,
  title={On the structure of the atom: an investigation of the stability and periods of osciletion of a number of corpuscles arranged at equal intervals around the circumference of a circle; with application of the results to the theory atomic structure},
  author={Tomson, JJ},
  journal={Philos. Mag. Series},
  volume={6},
  number={7},
  pages={237},
  year={1904}
}

@article{de2000spectroscopic,
  title={Spectroscopic properties and design of highly luminescent lanthanide coordination complexes},
  author={De Sa, GF and Malta, Oscar L and de Mello Doneg{\'a}, C and Simas, AM and Longo, RL and Santa-Cruz, PA and da Silva Jr, EF},
  journal={Coordination Chemistry Reviews},
  volume={196},
  number={1},
  pages={165--195},
  year={2000},
  publisher={Elsevier}
}

@article{lermusiaux2021toward,
  title={Toward Huygens’ sources with dodecahedral plasmonic clusters},
  author={Lermusiaux, Laurent and Many, V{\'e}ronique and Barois, Philippe and Ponsinet, Virginie and Ravaine, Serge and Duguet, Etienne and Tr{\'e}guer-Delapierre, Mona and Baron, Alexandre},
  journal={Nano Letters},
  volume={21},
  number={5},
  pages={2046--2052},
  year={2021},
  publisher={ACS Publications}
}

@article{roach2023symmetric,
  title={Symmetric plasmonic nanoparticle clusters: Synthesis and novel optical properties},
  author={Roach, Lucien and Lermusiaux, Laurent and Baron, Alexandre and Tr{\'e}guer-Delapierre, Mona},
  journal={Encyclopedia of Nanomaterials},
  pages={113--127},
  year={2023},
  publisher={Elsevier}
}

@article{many2019high,
  title={High optical magnetism of dodecahedral plasmonic meta-atoms},
  author={Many, V{\'e}ronique and D{\'e}zert, Romain and Duguet, Etienne and Baron, Alexandre and Jangid, Vikas and Ponsinet, Virginie and Ravaine, Serge and Richetti, Philippe and Barois, Philippe and Tr{\'e}guer-Delapierre, Mona},
  journal={Nanophotonics},
  volume={8},
  number={4},
  pages={549--558},
  year={2019},
  publisher={De Gruyter}
}

@article{Wales2006, 
year = {2006}, 
title = {{Structure and dynamics of spherical crystals characterized for the Thomson problem}}, 
author = {Wales, David J. and Ulker, Sidika}, 
journal = {Physical Review B}, 
issn = {1098-0121}, 
doi = {10.1103/physrevb.74.212101}, 
pages = {212101}, 
number = {21}, 
volume = {74}, 
}

@article{Sanz-Paz2006, 
year = {2018}, 
title = {{Enhancing Magnetic Light Emission with All-Dielectric Optical Nanoantennas}}, 
author = {Sanz-Paz, Maria and Ernandes, Cyrine and Esparza, Juan Uriel and Burr, Geoffrey W. and Hulst, Niek F. van and Maitre, Agnès and Aigouy, Lionel and Gacoin, Thierry and Bonod, Nicolas and Garcia-Parajo, Maria F. and Bidault, Sébastien and Mivelle, Mathieu}, 
journal = {Nano Letters}, 
issn = {1530-6984}, 
doi = {10.1021/acs.nanolett.8b00548}, 
pmid = {29701991}, 
pages = {3481--3487}, 
number = {6}, 
volume = {18}, 
}

@book{Cowan1981, 
year = {1981}, 
title = {{The Theory of Atomic Structure and Spectra}}, 
author = {Cowan, Robert D.}, 
isbn = {9780520038219}, 
publisher = {University of California Press}, 
edition = {1st}
}

@article{Purcell1946, 
year = {1946}, 
title = {{Spontaneous Emission Probabilities at Radio Frequencies}}, 
author = {Purcell, E. M.}, 
journal = {Physical Review}, 
issn = {0031-899X}, 
doi = {10.1103/physrev.69.674}, 
volume = {69}
}

@article{Karaveli2013, 
year = {2013}, 
title = {{Time-Resolved Energy-Momentum Spectroscopy of Electric and Magnetic Dipole Transitions in Cr3+:MgO}}, 
author = {Karaveli, Sinan and Wang, Shutong and Xiao, Gang and Zia, Rashid}, 
journal = {ACS Nano}, 
issn = {1936-0851}, 
doi = {10.1021/nn402568d}, 
pmid = {23879390}, 
pages = {7165--7172}, 
number = {8}, 
volume = {7}, 
}

@article{Vaskin2019, 
year = {2019}, 
title = {{Manipulation of Magnetic Dipole Emission from Eu3+ with Mie-Resonant Dielectric Metasurfaces}}, 
author = {Vaskin, Aleksandr and Mashhadi, Soheila and Steinert, Michael and Chong, Katie E. and Keene, David and Nanz, Stefan and Abass, Aimi and Rusak, Evgenia and Choi, Duk-Yong and Fernandez-Corbaton, Ivan and Pertsch, Thomas and Rockstuhl, Carsten and Noginov, Mikhail A. and Kivshar, Yuri S. and Neshev, Dragomir N. and Noginova, Natalia and Staude, Isabelle}, 
journal = {Nano Letters}, 
issn = {1530-6984}, 
doi = {10.1021/acs.nanolett.8b04268}, 
pmid = {30605616}, 
pages = {1015--1022}, 
number = {2}, 
volume = {19}, 
}

@article{Choi2016, 
year = {2016}, 
title = {{Selective Plasmonic Enhancement of Electric- and Magnetic-Dipole Radiations of Er Ions}}, 
author = {Choi, Bongseok and Iwanaga, Masanobu and Sugimoto, Yoshimasa and Sakoda, Kazuaki and Miyazaki, Hideki T.}, 
journal = {Nano Letters}, 
issn = {1530-6984}, 
doi = {10.1021/acs.nanolett.6b02200}, 
pmid = {27436631}, 
pages = {5191--5196}, 
number = {8}, 
volume = {16}, 
}

@article{Hein2013, 
year = {2013}, 
title = {{Tailoring Magnetic Dipole Emission with Plasmonic Split-Ring Resonators}}, 
author = {Hein, Sven M. and Giessen, Harald}, 
journal = {Physical Review Letters}, 
issn = {0031-9007}, 
doi = {10.1103/physrevlett.111.026803}, 
pmid = {23889429}, 
pages = {026803}, 
number = {2}, 
volume = {111}, 
}

@article{Feng2018, 
year = {2018}, 
title = {{Isotropic Magnetic Purcell Effect}}, 
author = {Feng, Tianhua and Zhang, Wei and Liang, Zixian and Xu, Yi and Miroshnichenko, Andrey E.}, 
journal = {ACS Photonics}, 
issn = {2330-4022}, 
doi = {10.1021/acsphotonics.7b01016}, 
pages = {678--683}, 
number = {3}, 
volume = {5}, 
}

@article{feng2017ideal,
  title={Ideal magnetic dipole scattering},
  author={Feng, Tianhua and Xu, Yi and Zhang, Wei and Miroshnichenko, Andrey E},
  journal={Physical review letters},
  volume={118},
  number={17},
  pages={173901},
  year={2017},
  publisher={APS}
}

@article{Sugimoto2021, 
year = {2021}, 
title = {{Magnetic Purcell Enhancement by Magnetic Quadrupole Resonance of Dielectric Nanosphere Antenna}}, 
author = {Sugimoto, Hiroshi and Fujii, Minoru}, 
journal = {ACS Photonics}, 
issn = {2330-4022}, 
doi = {10.1021/acsphotonics.1c00375}, 
pages = {1794--1800}, 
number = {6}, 
volume = {8}, 
}

@article{Moroz2005, 
year = {2005}, 
title = {{A recursive transfer-matrix solution for a dipole radiating inside and outside a stratified sphere}}, 
author = {Moroz, Alexander}, 
journal = {Annals of Physics}, 
issn = {0003-4916}, 
doi = {10.1016/j.aop.2004.07.002}, 
pages = {352--418}, 
number = {2}, 
volume = {315}, 
}

@book{Taflove2005, 
year = {2005}, 
title = {{Computational Electrodynamics: the finite-difference time-domain method}}, 
author = {Taflove, Allen and Hagness, Susan C}, 
isbn = {9780121709600}, 
publisher = {Artech House}, 
edition = {3rd}
}

@book{Monk2003, 
year = {2003}, 
title = {{Finite Element Methods for Maxwell's Equations}}, 
author = {Monk, Peter}, 
isbn = {9780191708633}, 
publisher = {Oxford University Press}, 
doi = {10.1093/acprof:oso/9780198508885.001.0001}, 
month = {4}
}

@article{Lalanne2018, 
year = {2018}, 
title = {{Light Interaction with Photonic and Plasmonic Resonances}}, 
author = {Lalanne, Philippe and Yan, Wei and Vynck, Kevin and Sauvan, Christophe and Hugonin, Jean‐Paul}, 
journal = {Laser \& Photonics Reviews}, 
issn = {1863-8880}, 
doi = {10.1002/lpor.201700113}, 
eprint = {1705.02433}, 
number = {5}, 
volume = {12}, 
}

@article{Wu2023, 
year = {2023}, 
title = {{Modal analysis of electromagnetic resonators: User guide for the MAN program}}, 
author = {Wu, Tong and Arrivault, Denis and Yan, Wei and Lalanne, Philippe}, 
journal = {Computer Physics Communications}, 
issn = {0010-4655}, 
doi = {10.1016/j.cpc.2022.108627}, 
eprint = {2206.13886}, 
pages = {108627}, 
volume = {284}, 
}

@article{Sauvan2022, 
year = {2022}, 
title = {{Normalization, orthogonality, and completeness of quasinormal modes of open systems: the case of electromagnetism [Invited]}}, 
author = {Sauvan, Christophe and Wu, Tong and Zarouf, Rachid and Muljarov, Egor A and Lalanne, Philippe}, 
journal = {Optics Express}, 
doi = {10.1364/oe.443656}, 
pmid = {35299463}, 
eprint = {2112.08103}, 
pages = {6846}, 
number = {5}, 
volume = {30}, 
}

@article{Sauvan2013, 
year = {2013}, 
title = {{Theory of the Spontaneous Optical Emission of Nanosize Photonic and Plasmon Resonators}}, 
author = {Sauvan, C. and Hugonin, J. P. and Maksymov, I. S. and Lalanne, P.}, 
journal = {Physical Review Letters}, 
issn = {0031-9007}, 
doi = {10.1103/physrevlett.110.237401}, 
pmid = {25167528}, 
pages = {237401}, 
number = {23}, 
volume = {110}, 
}

@misc{note:SM,
  note = {See Supplemental Material at [link] for additional data concerning equivalent polyhedra and symmetry classification, spectral proprties of the N-pods, details on optimization and on the electric dipole mode of the dodecapod.}
}

@article{reynier2025nearfield,
  title={Nearfield control over magnetic light-matter interactions},
  author={Reynier, Beno{\^\i}t and Charron, Eric and Markovic, Obren and Gallas, Bruno and Ferrier, Alban and Bidault, S{\'e}bastien and Mivelle, Mathieu},
  journal={Light: Science \& Applications},
  volume={14},
  number={1},
  pages={127},
  year={2025},
  publisher={Nature Publishing Group UK London}
}

@article{brule2022magnetic,
  title={Magnetic and electric Purcell factor control through geometry optimization of high index dielectric nanostructures},
  author={Br{\^u}l{\'e}, Yoann and Wiecha, Peter and Cuche, Aur{\'e}lien and Paillard, Vincent and Colas des Francs, G{\'e}rard},
  journal={Optics Express},
  volume={30},
  number={12},
  pages={20360--20372},
  year={2022},
  publisher={Optica Publishing Group}
}

@article{utyushev2024generation,
  title={Generation of nearly pure and highly directional magnetic light in fluorescence of rare-earth ions},
  author={Utyushev, Anton D and Gaponenko, Roman and Sun, Song and Shcherbakov, Alexey A and Moroz, Alexander and Rasskazov, Ilia L},
  journal={Physical Review B},
  volume={109},
  number={4},
  pages={045413},
  year={2024},
  publisher={APS}
}

@article{Thill2012, 
year = {2012}, 
title = {{Spheres Growing on a Sphere: A Model to Predict the Morphology Yields of Colloidal Molecules Obtained through a Heterogeneous Nucleation Route}}, 
author = {Thill, Antoine and D\'{e}sert, Anthony and Fouilloux, Sarah and Taveau, Jean-Christophe and Lambert, Olivier and Lansalot, Muriel and Bourgeat-Lami, Elodie and Spalla, Olivier and Belloni, Luc and Ravaine, Serge and Duguet, Etienne}, 
journal = {Langmuir}, 
issn = {0743-7463}, 
doi = {10.1021/la301857h}, 
pmid = {22775494}, 
pages = {11575--11583}, 
number = {31}, 
volume = {28}, 
}

@article{johnson1972optical,
  title={Optical constants of the noble metals},
  author={Johnson, Peter B and Christy, R-WJPrB},
  journal={Physical review B},
  volume={6},
  number={12},
  pages={4370},
  year={1972},
  publisher={APS}
}

@article{Wu2019, 
 title={Strong Purcell Effect for Terahertz Magnetic Dipole Emission with Spoof Plasmonic Structure}, 
 volume={2}, 
 ISSN={2574-0970}, 
 DOI={10.1021/acsanm.8b02318}, 
 number={2}, 
 journal={ACS Applied Nano Materials}, 
 author={Wu, Hong-Wei and Li, Yang and Chen, Hua-Jun and Sheng, Zong-Qiang and Jing, Hao and Fan, Ren-Hao and Peng, Ru-Wen}, 
 year={2019},
 pages={1045–1052} }

@article{Puente2023, 
 title={Magnetic Purcell Enhancement in a Nanoantenna-Spherical Bragg Resonator Coupled System}, 
 volume={535}, 
 ISSN={0003-3804}, 
 DOI={10.1002/andp.202300147},
 number={11}, 
 journal={Annalen der Physik}, 
 author={García-Puente, Yalina and Kashyap, Raman},
 year={2023} }

@article{Wu_2021,
 title={Nanoscale Light Confinement: the Q’s and V’s}, 
 volume={8}, 
 ISSN={2330-4022}, 
 DOI={10.1021/acsphotonics.1c00336}, 
 number={6}, 
 journal={ACS Photonics}, 
 author={Wu, Tong and Gurioli, Massimo and Lalanne, Philippe}, 
 year={2021}, 
 pages={1522–1538} }

@article{xiong2020,
  title={On the constraints of electromagnetic multipoles for symmetric scatterers: eigenmode analysis},
  author={Xiong, Zhongfei and Yang, Qingdong and Chen, Weijin and Wang, Zhuoran and Xu, Jing and Liu, Wei and Chen, Yuntian},
  journal={Optics Express},
  volume={28},
  number={3},
  pages={3073--3085},
  year={2020},
  publisher={Optical Society of America}
}

@article{gladyshev2020,
  title={Symmetry analysis and multipole classification of eigenmodes in electromagnetic resonators for engineering their optical properties},
  author={Gladyshev, Sergey and Frizyuk, Kristina and Bogdanov, Andrey},
  journal={Physical Review B},
  volume={102},
  number={7},
  pages={075103},
  year={2020},
  publisher={APS}
}

\end{document}